\documentstyle[12pt]{article}

\begin{document}
\begin{flushright}
USITP-98-04 \\
OSLO-TP 2-98 \\
May 98
\end{flushright}
\bigskip\bigskip
\Large
\begin{center}
\bf{Some New Non-Abelian 2D Dualities}\\

\bigskip

\normalsize
\bigskip

H. Gustafsson$^{a,}$\footnote{e-mail: henrik@physto.se},
S.E.  Hjelmeland$^{b,}$\footnote{e-mail: s.e.hjelmeland@fys.uio.no},
U. Lindstr\"om$^{a,b,}$\footnote{e-mail: ul@physto.se
and ulfl@boson.fys.no}
\bigskip \\
$^{a}${\it  ITP, University of Stockholm\\
Box 6730, Vanadisv\"agen 9, S-113 85 Stockholm\\
SWEDEN}\\
\end{center}
\begin{center}
\normalsize
{and}
\end{center}
\normalsize
\begin{center}
                $^{b}${\it Institute of Physics, University of Oslo\\
                \it P.O. BOX 1048, N-0316 Blindern, Oslo 3,\\
                \it NORWAY}
\end{center}
\vspace{3.0cm}
\normalsize

{\bf Abstract:} Starting from certain $3D$ non-abelian 
dual systems, we discuss
a number of related dual systems in $2D$, 
some of which are obtained by
dimensional reduction. 
The dualities relate massive scalar and vector fields,
and may be relevant for string theory in the context of
massive type IIA supergravity.
Supersymmetric extensions of the models are also presented.
\bigskip\bigskip\bigskip

\thispagestyle{empty}\eject
\setcounter{page}{1}

\section{Introduction}

Duality between different description of one and the same physical
system has a long history \cite{pamd}. Recently string theory has
focused the
interest on two dimensional scalar-scalar 
duality \cite{amit}, where also
non-abelian
duality has played a role \cite{ossa}. 
Non-abelian generalizations of
other types of
duality has been hard to come by, 
although there are some examples, such as the
$3D$ non-abelian vector-vector duality of \cite{Kar87a,Kar87b}. 
In this paper we
take these latter results as our 
starting point for an investigation of some
non-abelian dualities in $2D$. The non-abelian dualities we present
relate vectors to vectors in $3D$ and massive scalar
and
vector models in $2D$. We envisage possible applications to string
theory
in the context of massive type
IIA supergravity \cite{Rom,Ber}. {\it E.g.}, the 
massive D2-brane has been
shown to be dual to a dimensional reduction of the M2-brane coupled
to an auxiliary vector field \cite{Loz} via a vector-vector duality. 
Our results would be relevant if one wanted to extend these
considerations to include non-abelian fields and/or go to strings
using double dimensional reduction.

The usual $2D$ non-abelian duality has a geometrical interpretation
and generalization in terms of Poisson-Lie duality \cite{Kli95}.
We believe that our $2D$-results cannot be framed in that
language. 

It is also hoped that the present results contribute to the
understanding of the rich structure of $2D$ field theory.

We have chosen to present the models in terms of scalar and vector
fields. We
could have made use of the local $2D$ equivalence
$$
A_a=\partial_a\varphi + \epsilon_a^{\ b}\partial_b\chi
$$
to write the actions in terms of scalar fields only.
Barring nonlocal field redefinitions, these actions would be
higher order in derivatives, however.

The plan of the paper is as follows: In section \ref{sec:notation}
we introduce a short-hand
notation for duality systems that will facilitate the presentation
later. In
section \ref{sec:3D} we give a quick review of the
$3D$ results in \cite{Kar87a,Kar87b}. Section \ref{sec:2Dnew} 
contains new $2D$ dualities
that have no $3D$ counterpart, along with their
supersymmetric versions. In section
\ref{sec:reduction} we give the direct dimensional
reduction
of the $3D$ results collected in section \ref{sec:3D}. 
This is followed by some brief
conclusions and an appendix containing 
our notation and conventions as well as
some of the more cumbersome expressions for $2D$ actions.


\section{Notation}
\label{sec:notation}

\begin{figure}
\begin{center}
\setlength{\unitlength}{1mm}
\begin{picture}(100,70)
\put(43,60){$\bf{S(A,B)}$}
\put(46,57){\vector(-2,-3){23}}\put(54,57){\vector(2,-3){23}}
\put(26,40){$\bf{\delta B}$}
\put(68,40){$\bf{\delta A}$}
\put(18,17){$\bf{S(A)}$}
\put(78,17){$\bf{S(B)}$}
\put(20,10){\vector(0,1){5}}\put(80,10){\vector(0,1){5}}
\put(20,10){\line(1,0){20}}\put(60,10){\line(1,0){20}}
\put(46,9){\bf{dual}}
\end{picture}
\caption{\small The parent action $S(A,B)$ is (classically)
equivalent to both $S(A)$ and $S(B)$ showing
that $S(A)$ and $S(B)$ are dual to each other.}
\label{dual1-1}
\end{center}
\end{figure}
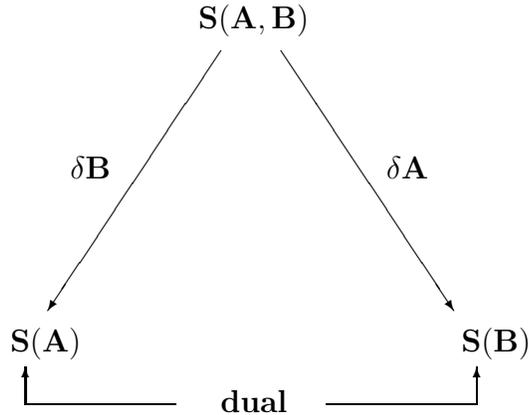

In what follows we will present a number of equivalences between
different
models, abelian, non-abelian and supersymmetric in two and three
dimensions.
To make the presentation as clear as possible 
we introduce the following
succinct notation: $S(A)$ denotes an action with an abelian symmetry
group
acting on the field $A$ in the argument. 
The action for a non-abelian theory
is denoted $I(A)$ and supersymmetry is indicated by bold face, {\it
e.g.}
${\bf S}(\Gamma)$ is an abelian 
supersymmetric action for the superfield
$\Gamma$. We further introduce the 
notation $(S(A),S(B))[S(A,B)]$ to
denote a {\em duality system} $(\sf{DS})$, {\it i.e.} the
dual pair of (abelian) actions
$S(A)$ and $S(B)$ with parent action $S(A,B)$
as illustrated in Fig. \ref{dual1-1}.
This notation will also
be used when one or two of the entries do not exist or are not known.
{\it E.g.}
$(S(A),S(B))[S(A,B)]\rightarrow (I(A),~\cdot ~)[~\cdot ~]$
means that $S(A)$ has a non-abelian extension $I(A)$, but that its
parent action and dual are lacking.

We have collected the rest of our notation and conventions in
Appendix \ref{app}.


\section{3D Vector-Vector Duality}
\label{sec:3D}

In this section we collect some known $3D$ results.
The primary actions are
\begin{eqnarray}
\label{eq:IBV3D}
 I_1(B,V)&=&\mbox{Tr}\int d^3x\left\{ \frac{1}{2} m^2 B_a B^{a}-
  \frac{1}{4} m\epsilon^{abc}B_a
  \left[ \nabla_b(V)B_c\right] \right\}+I(V),  \\
\label{eq:IA3D}
  I_1(A)&=&\mbox{Tr}\int d^3x \left\{-\frac{1}{4} F_{ab}F^{ab}+
  \frac{1}{2} m\Omega(A)\right\},\\
\label{eq:IAB3D}
   I(A,B)&=&\mbox{Tr}\int d^3x \left\{ \frac{1}{2} m^2 B_aB^a
        +\frac{1}{2} m\epsilon^{abc}
  B_aF_{bc}(A) +\frac{1}{2} m\Omega(A) \right\},
\end{eqnarray}
where $A$, $B$ and $V$ are fields in the adjoint representation of
some group $G$,
$F_{ab}(A)$ and $\Omega(A)$ denotes a field strength and a 
Chern-Simons
term, respectively, $\nabla_{b}(V)$ is a covariant derivative
and $I(V)$ is some action for the spectator field $V$.
The action $I_1(B,V)$ was first discussed in \cite{Tow84} and its
abelian version $I_1(B,V)\rightarrow S(B,V)$ is the self-dual model.
The action $I_1(A)$ was first presented in \cite{Des82}
and its abelian version
$I_1(A)\rightarrow S(A)$ is the topologically massive model.
With $I(A,B)\rightarrow S(A,B)$ and $V=0$, the $\sf{DS}$
$(S(A),S(B))[S(A,B)]$ was displayed in \cite{Des84}.

The non-abelian generalization of this was
subsequently investigated in \cite{Kar87a}
and was extended to include supersymmetry in \cite{Kar87b}. 
To briefly
recapitulate these developments we also need a second non-abelian
generalization of $S(B,V)$ namely
\begin{equation}
I_{2}(B,V)=I_{1}(B,V)
-\frac{m}{12}\mbox{Tr}\int d^3 x\epsilon^{abc}B_{a}B_{b}B_{c}.
\end{equation}

In \cite{Kar87a} the $\sf{DS}$ $(I_2(A,V),I_2(B,V))[I_2(\bar{A},B,V)]$
was derived.
Here the parent action is
\begin{eqnarray}
\nonumber
I_2(\bar{A},B,V)&=&\frac{1}{2} \mbox{Tr}\int d^3x \left\{m^2B_{a}B^{a}
 -\frac{1}{2} m\epsilon^{abc}B_a\left[ \nabla_b (V)B_c +\frac{1}{3}
B_bB_c\right]
 \right. \\
&&\qquad\qquad + \left.  m\Omega(\bar{A})\right\} +I(V).
\end{eqnarray}
It is invariant under two sets of gauge transformations
with parameters $\Lambda$ and $\Sigma$ respectively,
\begin{eqnarray}
\label{eq:I2AbarBVtransf}
& &(i)\ \ \ \delta_{\Lambda}B_{a}=[B_{a},\Lambda],
\ \ \ \delta_{\Lambda}V_{a}=
\nabla_{a}(V)\Lambda,\ \ \ \delta_{\Lambda}\bar{A}_{a}=0,
\nonumber \\ \nonumber \\
& &(ii)\ \ \ \delta_{\Sigma}B_{a}=0,\ \ \ \delta_{\Sigma}V_{a}=0,
\ \ \
\delta_{\Sigma}\bar{A}_{a}=\nabla_{a}(\bar{A})\Sigma.
\end{eqnarray}
The model $I_2(A,V)$ dual to $I_2(B,V)$ is
\begin{eqnarray}\nonumber
I_2(A,V)&=&\frac{1}{2} \int d^3x\left\{ 
-\frac{1}{4} {}^{\ast}F^{aA}(A)
 G_{aAbB}(A-V){}^{\ast} F^{bB}(A)+m\Omega(A) \right\}\\
&&\qquad \qquad +I(V),
\end{eqnarray}
where
\begin{equation}
(G^{-1})^{aAbB}\equiv\left[ \eta^{ab}\delta^{AB}+
(1/2m)\epsilon^{abc}(A-V)^C_c f^{ABC}\right].
\end{equation}
Here $A,B,...$ are Lie-algebra indices and $f^{ABC}$ are the 
structure constants.
The action $I_2(A,V)$ is invariant under
\begin{eqnarray}
& &(i)\ \ \ \delta_{\Lambda}A_{a}=-\frac{1}{2}[B_{a},\Lambda],\ \
\
\delta_{\Lambda}V_{a}=\nabla_{a}(V)\Lambda, \nonumber \\ \nonumber \\
& &(ii)\ \ \ \delta_{\Sigma}A_{a}=
\bar{\nabla}_{a}\Sigma\equiv\nabla_{a}(A)\Sigma
+\frac{1}{2}[B_{a},\Sigma],\ \ \ \delta_{\Sigma}V_{a}=0,
\end{eqnarray}
where, {\it e.g.}
\begin{eqnarray}
  [B_a,\Lambda ]^A=-\frac{1}{2m}\epsilon^{bde}F^E_{de}
  G_{bEaB}\Lambda_D f^{BDA}.
\end{eqnarray}
This duality is somewhat unusual in that it also involves a shift
$\bar{A} = A+\frac{1}{2}B$ before eliminating $B$ to get $I_2(A,V)$.

The introduction of the modified $I_2(B,V)$ was seen in \cite{Kar87a}
to be necessary for a straightforward dualization. The abelian
$\sf{DS}$
$(S(B),S(A))[S(A,B)]$ has two direct non-abelian (partial)
generalizations:
$(\tilde{I}(B),I_1(A))[I(A,B)]$ and
\mbox{$(I_1(B,V),~\cdot ~)[~\cdot ~]$.} More explicitly,
the non-abelian topologically
massive model of \cite{Des82} has $I(A,B)$ as parent 
action which leads
to dual action $\tilde{I}(B)$. However $\tilde{I}(B)$ is non-local
(it is found in a Fock-Schwinger gauge), and is not equivalent to
$I_1(B,V)$.
For the action $I_1(B,V)$ of \cite{Tow84}, on the other hand, 
no parent
action
and dual could be found.

All these results have their $N=1$ supersymmetric counterpart.
To state them we first give the supersymmetrization of the various
actions
found by letting
$I_{1}(B,V)\rightarrow{{\bf I}(\Psi,V)}$,
$I_{1}(A)\rightarrow{{\bf I}_{1}(\Gamma)}$,
$I(A,B)\rightarrow{{\bf I}(\Psi,\Gamma)}$,
$I_{2}(B,V)\rightarrow{{\bf I}_{2}(\Psi,V)}$,
$I_{2}(\bar{A},B,V)\rightarrow{{\bf I}_{2}(\bar{\Gamma},\Psi,V)}$ 
where
\begin{eqnarray}
{{\bf I}_1(\Psi,V)}&=&
-\frac{1}{2}\mbox{Tr}\int d^{3}xd^{2}\theta
\left[m^{2}\Psi^{\alpha}\Psi_{\alpha}
+\frac{1}{4}m
\Psi^{\beta}\nabla^{\alpha}\nabla_{\beta}\Psi_{\alpha}\right]
+I(V), \\
{{\bf I}_1(\Gamma)}&=&
\mbox{Tr}\int d^{3}x d^{2}\theta\left[ \frac{1}{2}W^{\alpha}W_{\alpha}
+m\Omega(\Gamma)\right] , \\
\label{PsiGamma}
{{\bf I}(\Psi,\Gamma)}&=&
\frac{1}{2}\mbox{Tr}\int d^{3}xd^{2}\theta
\left[ -m^2\Psi^{\alpha}\Psi_{\alpha}
+2m\left(\Psi^{\alpha}W_{\alpha}(\Gamma)+\Omega(\Gamma)\right)\right],
\\
{{\bf I}_{2}(\Psi,V)}&=& {\bf I}_1(\Psi,V)
-\frac{m}{24}\mbox{Tr}\int d^{3}xd^{2}\theta
\, \nabla^{\alpha}\Psi^{\beta}\{\Psi_{\alpha},\Psi_{\beta}\} , \\
{{\bf I}_{2}(\bar{\Gamma},\Psi,V)}&=& {\bf I}_2(\Psi,V)
+\mbox{Tr}\int d^{3}xd^{2}\theta
\, m\Omega(\bar{\Gamma}).
\end{eqnarray}
Here $\Psi$, $V$ and $\Gamma$ are spinorial superfields
in the adjoint representation of the gauge group $G$,
$W_{\alpha}$ the superfield strength, $\Omega$ the super
Chern-Simons term and $\nabla_{\alpha}$
super Yang-Mills covariant derivatives.
There is also an action ${\bf I}_2(\Gamma,V)$, the supersymmetric
generalization of $I_2(A,V)$, which we shall not need.
The results of \cite{Kar87b} are as follows.
First one has the supersymmetric abelian
$\sf{DS}$ $({\bf S}(\Psi),{\bf S}(\Gamma))[{\bf S}(\Psi,\Gamma)]$.
The corresponding non-abelian systems are
\mbox{$({\bf I}_{2}(\Psi,V),{\bf I}_{2}(\Gamma,V))[{\bf
I}_{2}(\bar{\Gamma},\Psi,V)]$}
and
\mbox{$(\tilde{\bf{I}},{\bf I}_{1}(\Gamma))[{\bf I}(\Psi,\Gamma)]$}
where
$\tilde{\bf{I}}$ is a non local supersymmetric extension of
$\tilde{I}(B)$. It is not equivalent to
${\bf I}_{1}(\Psi,V)$ and furthermore had to be found
by abandoning superspace and calculating in components. 
Finally again
$({\bf I}_{1}(\Psi,V),~\cdot ~)[~\cdot ~]$ was not 
possible to complete.

This concludes our summary of the $3D$ results which 
we collect in table
\ref{tab:DS3D}.

\begin{table}
\begin{center}
\begin{tabular}{|l||l|l|}   \hline
       { }     & {\bf Bosonic} &   {\bf Supersymmetric}  \\
           \hline \hline
             & & \\
 {\bf Abelian}  & ${\left( \begin{array}{c} S(B) \\ S(A) \end{array}
  \right)}_{S(A,B)}$
            & ${\left( \begin{array}{c} {\bf S}(\Psi) \\
               {\bf S}(\Gamma) \\
    \end{array}\right)}_{{\bf S}(\Psi,\Gamma)}$ \\
             & & \\ \hline & & \\
         & ${\left( \begin{array}{c} I_2(B,V) \\ I_2(A,V) \end{array}
  \right)}_{I_2(\bar{A},B,V)}$
            & ${\left( \begin{array}{c} {\bf I}_2(\Psi,V) \\
               {\bf I}_2(\Gamma,V) \\
  \end{array}\right)}_{{\bf I}_2(\bar{\Psi},\Gamma,V)}$ \\
                &  & \\
{\bf Non-Abelian}          & ${\left( \begin{array}{c}
          \mbox{{\it non-local}} \\ I_1(A) \end{array}
  \right)}_{I(A,V)}$
            & ${\left( \begin{array}{c} \mbox{{\it non-local}} \\
               {\bf I}_1(\Gamma) \\
  \end{array}\right)}_{{\bf I}_1(\Psi,\Gamma)}$ \\
             &  & \\
     & ${\left( \begin{array}{c} I_1(B,V) \\ \cdot \end{array}
  \right)}$
            & ${\left( \begin{array}{c} {\bf I}_1(\Psi,V) \\
               \cdot \\
  \end{array}\right)}$ \\
             & & \\ \hline
\end{tabular}
\caption{The duality systems in 3D.}
\label{tab:DS3D}
\end{center}
\end{table}


\section{New 2D Non-Abelian Duality}
\label{sec:2Dnew}

In this section we present a new non-abelian vector-vector duality in
$2D$
 derived from the $3D$
action  (\ref{eq:IAB3D}).  Dimensionally reduced this action becomes
\begin{eqnarray}
\label{AaBb}\nonumber
I(A,a,B,b)&=&\mbox{Tr}\int d^2x \left\{ \frac{1}{2} m^2
B_aB^a-\frac{1}{2}
m^2b^2 +  \frac{1}{2} mbF(A)        \right. \\
&+& \left. ma(F(A)+\epsilon^{ab}\nabla_a(A)B_b)
  \right\},
\end{eqnarray}
where the scalar fields $a$ and $b$ arise in dimensional reduction
according to
$A_a\to (A_a,a)$ and $B_a\to (B_a,b)$, and $F(A)\equiv
\epsilon^{ab}F_{ab}$.
As mentioned in the previous section, the $3D$ action is a 
parent action for
$I_1(A)$ and for a non-local $B$-action 
that is found after breaking gauge
invariance.  To find a local $B$-action one has 
to add an $\epsilon B B B$-term.
In $2D$ the situation is different. Without the $3D$-Lorentz 
invariance to
respect there is a larger freedom. 
We may choose to integrate out  both
$B_a$ and
$b$ to find $I(A,a)$, the dimensional reduction of $I_1(A)$:
\begin{eqnarray}
\label{Aa}
I(A,a)&=&\mbox{Tr}\int d^2x \left\{ -{1\over 4}F_{ab}F^{ab}+
\frac{1}{2}\nabla_a(A)a\nabla^a(A)a+maF(A)
  \right\},
\end{eqnarray}
or we may integrate out either $B_a$ or $b$ by itself yielding actions
which
we denote by $I(A,a,b)$ and
$I(A,a,B)$ respectively\footnote{In fact, $B_a$ is an auxiliary field,
so it might be most natural to always integrate it out. This can
always be done at any stage below.}.
Similarily, we may now contemplate integrating
out the vector $A_a$ by itself,
leaving the scalar $a$, {\it i.e.},
the third $3D$ vector component, in the action.
This can be done along the
lines of standard $2D$ non-abelian duality for scalar fields.
The $A_a$ field equations give
\begin{eqnarray}
A^A_a&=&{(M^{-1})}^{AB}h^B_a
\end{eqnarray}
where
\begin{eqnarray}
{(M^{-1})}^{AB}M^{BC}=\delta^{AC},\\
M^{AB}\equiv (\frac{1}{2} b +a)^Cf^{CAB},
\end{eqnarray}
and
\begin{eqnarray}
h^B_a\equiv\frac{1}{2} f^{BCD}a^CB^D_a-\partial_a(\frac{1}{2} b+a)^B
\end{eqnarray}
Here $A,B,..$ are again adjoint group indices and $f^{ABC}$ are the
Lie-algebra
structure  constants.
The corresponding action, dual to  $I(A,a,b)$ (or, equivalently to
$I(A,a)$), is;
\begin{eqnarray}
\label{Bba}
I(B,b,a)&=&\int d^2x \left\{ \frac{1}{2} m^2B_a^A B^{aA}
-\frac{1}{2} m^2 b^A b^A +m\epsilon^{ab}h^A_a{(M^{-1})}^{AB}h^B_b
 \right. \nonumber \\
&+& \left. m\epsilon^{ab}a^A\partial_a B^A_b \right\} .
\end{eqnarray}
Despite the non-covariant appearence, the action
(\ref{Bba}) is gauge invariant. It represents a theory with an
anti-symmetric
piece added to the group metric in
the $B^2$-term,
non-linear $\sigma$-model target-space torsion type  coupling for the
scalars
involving
also the vector $B$-field,
and non-linear scalar vector interaction terms. In agreement with the
$3D$
result, integrating out the
scalar $a$ to yield a local theory is not possible.

Having thus described the $2D$ $\sf{DS}$
$(I(A,a,b),I(B,b,a))[I(A,a,B,b)]$, we
now turn to its
supersymmetric version. In this subsection we write out the 
vector and spinor
indices explicitly, using
the ``$\pm$  notation'' introduced in Appendix \ref{app2D}.

The supersymmetrization of $I(A,a,B,b)$ is  
given by the $2D$ version of
${\bf I}(\Psi, \Gamma)$ in
(\ref{PsiGamma}),
\begin{eqnarray}\nonumber
{\bf I}(\Psi,\Gamma)&=&
\mbox{Tr}\int d^{2}xd^{2}\theta
\left\{im^2\Psi_+\Psi_-
+im\nabla_{(+}\Psi_{-)}H-2mH^2 \right\},
\end{eqnarray}
where again $\Psi$ is a ($2D$) spinor superfield and $H$ is the $\Gamma$
field-strength introduced
in Appendix  \ref{app2D}.  To see how the dualization works, we perform
the
$\theta$
integration and find the
component action. We define the component fields using projections. The
components of $\Gamma$ are, (in a Wess-Zumino gauge),
\begin{eqnarray}\nonumber
a &\equiv & \sqrt{2}H|, \quad  A_{+\!\!\!+ /=} \equiv
-iD_{\pm}\Gamma_{\pm}|
-\frac{1}{2} \{\Gamma_{\pm},\Gamma_{\pm}\}|, \\
\lambda_{\pm} &\equiv & W_{\pm}| = \pm \nabla_{\pm}H|, \quad F(A) \equiv
2\sqrt{2}\, i\nabla_+\nabla_-H|,
\end{eqnarray}
where $|$ denotes ``the $\theta$-independent part of''. The components
of
$\Psi$ are
\begin{eqnarray}\nonumber
\zeta_{\pm} &\equiv & \Psi_{\pm}|, \quad B_{+\!\!\!+/=} \equiv \mp i
\nabla_{\pm}\Psi_{\pm}|,
\quad b \equiv -\frac{i}{\sqrt{2}}\nabla_{(+}\Psi_{-)}|,\\
S &\equiv &-\frac{i}{\sqrt{2}}\nabla_{[+}\Psi_{-]}|, \quad
\chi_{\pm}\equiv
\mp\frac{i}{2}\nabla_{\pm}^2
\Psi_{\mp}| \pm\frac{i}{2}\nabla_{\mp}\nabla_{\pm}\Psi_{\pm}|.
\end{eqnarray}
Using these definitions we derive the following component action
\begin{eqnarray}
\label{eq:IGammaPsiComp}
\nonumber
&&{\bf I}(A,a,\lambda,B,b, \zeta,\chi,S)\\ \nonumber
&=&2\mbox{Tr}\int d^{2}x\left\{m^2\left( -B_{+\!\!\!+}
B_=-\frac{1}{2} b^2 +\frac{1}{2} S^2
+ i(\zeta_-D_{+\!\!\!+}\zeta_--\zeta_+D_=\zeta_+)
\right. \right. \\ \nonumber
&-&\left. \left.
2i\chi_{[-}\zeta_{+]}+\frac{1}{\sqrt{2}}[a,\zeta_{-}]\zeta_{+}
\right) +4im\lambda_{-}\lambda_{+}+maF(A)
\right. \\
&-& \left.
 \frac{ma}{\sqrt{2}}D_{[+\!\!\!+}B_{=]}-2im\chi_{[+}\lambda_{-]}
+\frac{1}{2}mbF(A) \right\}.
\end{eqnarray}
We recognize the bosonic action $I(A,a,B,b)$ in
(\ref{eq:IGammaPsiComp}),
 plus supersymmetric completions.
Integrating out the fields $B,b,\zeta,\chi$ and $S$ we get
\begin{eqnarray}
  {\bf I}(A,a,\lambda)=2I(A,a) &+& 2\mbox{Tr}\int d^2x
  \left\{\rule{0cm}{0.5cm}\lambda_{-}D_{+\!\!\!+}\lambda_{-}-
  \lambda_{+}D_{=}\lambda_{+} \right. \nonumber \\
  &+& \left. \frac{1}{\sqrt{2}}[a,\lambda_{-}]\lambda_{+}+
  4im\lambda_{-}\lambda_{+} \rule{0cm}{0.5cm}\right\} ,
\end{eqnarray}
whereas by varying $A$ and $\lambda$ we arrive at
\begin{eqnarray}
{\bf I}(a,B,b,\zeta,\chi,S)&=& 2I(B,b,a)+
2\mbox{Tr}\int d^2x \left\{\rule{0cm}{0.5cm} m^2
\left(\zeta_-D_{+\!\!\!+}\zeta_--\zeta_+D_=\zeta_+
 \right. \right.  \nonumber \\
&+& \left. \left. \frac{1}{\sqrt{2}}[a,\zeta_{-}]\zeta_{+}
-2i\chi_{[-}\zeta_{+]}+\frac{1}{2}S^2\right)
+im\chi_{+}\chi_{-} \right\} .
\end{eqnarray}
The dualities found in this section are presented
in table \ref{tab:DS2Dnew}.

\begin{table}
\begin{center}
\begin{tabular}{|l||l|l|}   \hline
       { }     & {\bf Bosonic} &   {\bf Supersymmetric}  \\
           \hline \hline
             & & \\
 {\bf Non-Abelian}    & ${\left( \begin{array}{c} I(A,a)
 \\ I(B,b,a) \end{array}
  \right)}_{I(A,a,B,b)}$
            & ${\left( \begin{array}{c} {\bf I}(A,a,\lambda) \\
               {\bf I}(a,B,b,\zeta,\chi,S) \\
    \end{array}\right)}_{{\bf I}(\Psi,\Gamma)}$ \\
                       & & \\ \hline
\end{tabular}
\caption{The new non-abelian dualities in 2D.}
\label{tab:DS2Dnew}
\end{center}
\end{table}


\section{3D (self-dual) non-abelian duality from 2D point of view}
\label{sec:reduction}

So far we have discussed dualities of the reduced theory that
have no correspondence in $3D$. However, we may also mimic the
three dimensional non-abelian dualization.
To this end we reduce  $I_2(\bar{A},B,V)$ to $2D$:
\begin{eqnarray}
\label{eq:I2AbarabarBbVv}
 I_2(\bar{A},\bar{a},B,b,V,v)=
 \mbox{Tr} \int d^2x \left\{\frac{1}{2} m^2 B_a B^a
- \frac{1}{2} m^2 b^2 \right. \nonumber \\ \left.
- \frac{1}{2} mb\epsilon^{ab} \left(
\nabla_a(V)B_b+\frac{1}{2} B_a B_b \right)
\right. \nonumber \\ \left.
    +\frac{1}{4}m\epsilon^{ab}B_a [v,B_b]
  +m\bar{a} F(\bar{A}) \right\}+I(V,v).
\end{eqnarray}
The action is invariant under two sets of gauge transformations
\begin{eqnarray}
& &(i) \ \ \delta_{\Lambda}B_a =[B_a,\Lambda],\ \ \
\ \
\delta_{\Lambda}b=[b,\Lambda], \nonumber \\
& &\ \ \ \ \ \ \delta_{\Lambda}V_a=\nabla_a (V)\Lambda,\ \
\ \ \
\delta_{\Lambda}v=[v,\Lambda], \nonumber \\
& &\ \ \ \ \ \ \delta_{\Lambda}\bar{A}_a=0,\ \ \ \ \
\delta_{\Lambda}\bar{a}=0. \nonumber\\
\nonumber \\
& &(ii) \ \ \delta_{\Sigma}B_a=0,\ \ \ \ \
\delta_{\Sigma}b=0, \nonumber \\
& &\ \ \ \ \ \ \delta_{\Sigma}V_a=0,\ \ \ \ \
\delta_{\Sigma}v=0, \nonumber \\
& &\ \ \ \ \ \ \delta_{\Sigma}\bar{A}_a=
\nabla_a (\bar{A})\Sigma ,\ \ \ \ \
\delta_{\Sigma}\bar{a}=[\bar{a},\Sigma]
\end{eqnarray}
which are the dimensional reduced version of
eq. (\ref{eq:I2AbarBVtransf}).

Variation with respect to $\bar{a}$ implies that $\bar{A}_a$ is
pure gauge and can be dropped from the action. We then get
an action $I_2(B,b,V,v)$ which is the $3D$ action $I_2(B,V)$
reduced to $2D$.
If we  shift $(\bar{A}_a,\bar{a})=(A_a,a)+
\frac{1}{2} (B_a,b)$ in (\ref{eq:I2AbarabarBbVv}) we get
the equivalent action,
\begin{eqnarray}
\label{eq:I2AaBbVv}
  I_2(A,a,B,b,V,v)=\mbox{Tr}
\int d^2x \left\{\frac{1}{2} m^2 B_a B^a
 - \frac{1}{2} m^2 b^2 +\frac{1}{2} mb F(A)
   \right. \nonumber \\  \left.
 + m\epsilon^{ab}B_a\nabla_b(A)a
+\frac{1}{2}m(a-v)\epsilon^{ab}B_a B_b
   \right. \nonumber \\  \left.
 +mb\epsilon^{ab}(A_a-V_a)B_b
 +ma F(A) \right\}+I(V,v).
\end{eqnarray}
From the equations of motions for $B_a$ and $b$ we
deduce
\begin{eqnarray}
  && B^{a A} G_{a A}^{-1 \ \ b B} =k^{b B}, \\
  && b^A=\frac{1}{2m}F^A(A)+\frac{1}{2m}\epsilon^{ab}
     (A_a^B-V_a^B )B_b^C f^{ABC},
\end{eqnarray}
where
\begin{eqnarray}
  && k^{a A}\equiv -\frac{1}{4m^2}\epsilon^{ab}
  F^C (A)\nabla_b a^A
  -\frac{1}{4m^2}\epsilon^{ab}(A^B_{b}-V^B_{b})
   F^C(A) f^{ABC}, \\
  && G_{aA bB}^{-1}\equiv\delta_{AB}\eta_{ab}
  +\frac{1}{2m}\epsilon_{ab}f_{ABC}(a^C-v^C)
  \nonumber \\
  && \ \ \ \ \  -\frac{1}{4m^2}\epsilon_{ac}\epsilon_{bd}
  (A^{c C}-V^{c C})(A^{d D}-V^{d D})
    f_A^{\ EC}f_B^{\ DE}.
\end{eqnarray}
Using these equations we can eliminate
$B$ and $b$ in (\ref{eq:I2AaBbVv}). We then
obtain an action dual to $I_2(B,b,V,v)$,
\begin{eqnarray}
\label{action:I2AaV}
 I_2(A,a,V,v)=\int d^2x \left\{-\frac{1}{2} m^2\, {}^{\ast}k^{aA}
G_{a A dD}(A-V,a-v) {}^{\ast} k^{dD}   \right. \nonumber \\ \left.
-\frac{1}{4} F_{ab}^A(A) F^{ab A}(A)
  +ma^A F^A(A)\right\} +I(V,v) ,
\end{eqnarray}
containing two different gauge fields. 
The gauge transformations are now
\begin{eqnarray}
&&(i) \ \ \delta_{\Lambda}A_{a}
=-\frac{1}{2}[B_{a},\Lambda],\ \ \ \ \
\delta_{\Lambda}a=-\frac{1}{2}[b,\Lambda], \nonumber \\
&&\ \ \ \ \ \ \delta_{\Lambda}V_{a}=\nabla_{a}(V)\Lambda,
\ \ \ \ \ \delta_{\Lambda}v=[v,\Lambda] ,
\nonumber \\ \nonumber \\
&&(ii) \ \ \delta_{\Sigma}A_{a}=
\bar{\nabla}_{a}\Sigma\equiv
\nabla_{a}(A)\Sigma+\frac{1}{2}[B_{a},\Sigma],
\nonumber \\
&&\ \ \ \ \ \ \delta_{\Sigma}a=[\bar{a},\Sigma]\equiv[a,\Sigma]
+\frac{1}{2}[b,\Sigma], \nonumber \\
&&\ \ \ \ \ \ \delta_{\Sigma}V_{a}=0,\ \ \ \ \
\delta_{\Sigma}v=0,
\end{eqnarray}
where, {\it e.g.}
\begin{eqnarray}
 [B^a,\Lambda ]^C=G^{aAdD}k_{dD}\Lambda^B f^{ABC}.
\end{eqnarray}
Thus we have the {\sf DS} $(I_2(A,a,V,v),I_2(B,b,V,v))
[I_2(\bar{A},\bar{a},B,b,V,v)]$ in analogy with the $3D$ result.


We now turn to the supersymmetric generalization of the self-dual
theory. The
procedure follows closely that of the bosonic theory above and is
essentially
the same as in
$3D$. The proliferation of fields and terms in the actions makes the
expressions
a bit difficult to appreciate, though, and for this reason we have
choosen to be
a bit sketchy below, emphasizing mainly the structure.

The two dimensional analogue of the first order action
${\bf I}_2(\bar{\Gamma},\Psi,V)$ is
\begin{eqnarray}
{\bf I}_2(\Psi ,\bar{H},V,v)&=& -\frac{1}{2}\mbox{Tr}\int d^2 xd^2
\theta
\left[2im^2\Psi_-\Psi_+
+\frac{m}{4}\nabla_{(+}\Psi_{-)}\nabla_{(+}\Psi_{-)}
\right. \nonumber \\
&+& \left. \frac{m}{2}\{ \Psi_+ ,\Psi_- \} \nabla_{(+}\Psi_{-)}
-\frac{m}{6}(\nabla_{-}\Psi_{-})\Psi_{+}^{2} \right. \nonumber \\
&-& \left. \frac{m}{6}(\nabla_{+}\Psi_{+})\Psi_{-}^{2}
+\frac{m}{2}\{\Psi_{+},\Psi_{-}\}{\cal V}+4m\bar{H}^2 \right]+ {\bf
I}(V,v)
\nonumber \\
\end{eqnarray}
where ${\cal V}$ is defind analogously to $H$ but with the difference
that the spinorial gauge field is $V$ instead of $\Gamma$.
Integrating out the Grassmann variables we find the component action
(52),
given in the appendix \ref{appActions}. A shift
$({\bar{A}}_a,\bar a) = (A_a,a)+\frac{1}{2}(B_a,b)$ allows us to
eliminate the $(B_a,b)$ multiplet and find a complicated system
consisting of the fields contained in the spinorial gauge multiplet
$\Gamma$
and the rest of the fields contained in the spinorial 
superfield $\Psi$
multiplet which was not integrated out.

\section{Conclusions}

In this paper we have studied the reduction to $2D$ of certain
$3D$ systems. We have seen that the $3D$ non-abelian dualities
reduce to $2D$ non-abelian dualities
between massive scalar and vector fields and we have presented
new non-abelian dualities in the $2D$ systems that have no
$3D$ counterpart. Further we have given the supersymmetrizations
of these relations.

This investigation arose out of a wish to see if the
$3D$ non-abelian dualities could lead to new types of
non-abelian $2D$ dualities, possibly extending the set of
$T$-dualities for strings.
It was hoped that the $2D$-equivalence
$A_a=\partial_a \varphi+\epsilon_a^{\ b}\partial_b\lambda$\,
in conjunction with nonlocal field redefinitions would allow
for such an application. The reduced systems turned out to be
much too unwieldly, however.
As they stand, the $2D$ dualities presented involve massive
scalar and vector fields and one has to look elsewhere for
applications. It is then interesting to note that there are other
extended objects within string/M-theory that do involve massive
dualities. In particular, the duality between a massive D2-brane and
a dimensionally reduced M2-brane coupled to an auxiliary
vector field exemplifies this \cite{Loz}. These models involve
Born-Infeld 
terms in place of our $B^2$-terms, but we may speculate that,
as usual, a non-abelian
generalization would start from the first term in a series
expansion. This would directly lead to some of our $3D$ actions. Our
$2D$-results would then be relevant for the double dimensional
reduction of
these models.

\bigskip
\begin{flushleft}
{\bf Acknowledgments:} We are grateful to Martin Ro\v{c}ek
and Rikard von Unge for reading and commenting the manuscript.
SEH would like to thank ITP at Stockholm for hospitality
during part of this research.
The research of UL was supported in part by NFR grant
No F-AA/FU 04038-312  and by NorFA grant No 9660003-0.
\bigskip
\end{flushleft}


\appendix

\section{Notation and Conventions}
\label{app}
$a,b,\ldots$ and $\alpha,\beta,\ldots$ denote 
Lorentz vector and spinor
indices, respectively. We do not discriminate between $3D$ and $2D$
indices
since they appear in different sections. In $D=2$ we will also use
$a\in\{+\!\!\!+,=\}$ and $\alpha\in\{+,-\}$. Our metric has signature
$(+,-,-)$ and $(+,-)$ in $D=3$ and $D=2$. We (anti)symmetrize without
combinatorial factors, {\it e.g.} $A_{(a}B_{b)}\equiv
A_{a}B_{b}+A_{b}B_{a}$.
All our fields $A=A^{B}T^{B}$ transform in the adjoint representation
of some Lie group
$G$ whose Lie algebra generators satisfy
\begin{equation}
\mbox{Tr}(T^{A}T^{B})=\delta^{AB},~~~~~~[T^{A},T^{B}]=f^{ABC}T^{C}
\end{equation}
Covariant derivatives act according to
\begin{equation}
(\nabla_a(A)B_b)^{A}=\partial_a B_b^{A}+f^{ABC}A_a^{B}B_b^{C}
\end{equation}
The field strength in $D=3$ is
$F_{ab}(A)=\partial_{a}A_{b}-\partial_{b}A_{a}+[A_{a},A_{b}]$
and in $D=2$ $F=\partial_{+\!\!\!+}A_{=}-\partial_{=}A_{+\!\!\!+}
+[A_{+\!\!\!+},A_{=}]$.
The dual field strength is ${}^\ast F^a\equiv\epsilon^{abc}F_{bc}$ in
3D and ${}^\ast F^a\equiv\epsilon^{ab}F_{b}$ in 2D.
The (3D) Chern-Simons term is
\begin{equation}
\Omega(A)=\epsilon^{abc}A_a\left( F_{bc}-
  \frac{2}{3}A_bA_c\right)=
2\epsilon^{abc}A_a\left(\partial_b A_c+\frac{2}{3}A_b A_c \right) .
\end{equation}

\subsection{3D-superspace}
\label{app3D}

In superspace we replace vector indices by pairs of spinor indices
according
to
\begin{equation}
V_{\alpha\beta}=\frac{1}{\sqrt{2}}(\gamma^{a})_{\alpha\beta}V_{a},
\end{equation}
where $\{(\gamma^{a})_{\alpha}^{~\beta},(\gamma^{b})_{\beta}^{~\nu}\}
=2\eta^{ab}\delta_{\alpha}^{~\nu}$.
We raise and lower indices using
$C_{\alpha\beta}=-C^{\alpha\beta}=\sigma^2$,
 {\it e.g.} $\Psi^{\alpha}=C^{\alpha\beta}\Psi_{\beta}$;
$\psi_{\alpha}=\Psi^{\beta}C_{\beta\alpha}$.
These are essentially the conventions of \cite{Gat83}.
Letting $D_{\alpha}$ be the flat superspace covariant derivatives,
the Yang-Mills covariant derivatives
$\nabla_{\alpha}=D_{\alpha}-i\Gamma_{\alpha}$
and
$\nabla_{\alpha\beta}=\partial_{\alpha\beta}-i\Gamma_{\alpha\beta}$
satisfy
\begin{eqnarray}
\{\nabla_{\alpha},\nabla_{\beta}\}&=&2i\nabla_{\alpha\beta}, \\
~[\nabla_{\alpha},\nabla_{\beta\gamma}]&=&C_{\alpha(\beta}W_{\gamma)},
\\
\{\nabla^{\alpha},W_{\alpha}\}&=&0 ,
\end{eqnarray}
where $W_{\alpha}$ is the super Yang-Mills spinorial field strength
$W_{\alpha}
=\frac{1}{2}D^{\beta}D_{\alpha}\Gamma_{\beta}
-\frac{i}{2}[\Gamma^{\beta},D_{\beta}\Gamma_{\alpha}]
-\frac{1}{6}[\Gamma^{\beta},\{\Gamma_{\beta},\Gamma_{\alpha}\}]$ with
$\Gamma_{\alpha\beta}
=-\frac{i}{2}D_{(\alpha}\Gamma_{\beta)}
-\frac{1}{2}\{\Gamma_{\alpha},\Gamma_{\beta}\}$.
The super Yang-Mills Chern-Simons
form corresponds
\begin{equation}
\Omega(\Gamma)=
\Gamma^{\alpha}(W_{\alpha}-\frac{1}{6}[\Gamma^{\beta},
\Gamma_{\alpha\beta}]).
\end{equation}
A convenient representation of the Clifford algbra is
$(\gamma^{a})_{\alpha}^{~\beta}
=(\sigma^{2},-i\sigma^{1},i\sigma^{3})$.

\subsection{2D-superspace}
\label{app2D}

We reduce to $2D$ by letting
$(\gamma^a)_{\alpha}^{~\beta}=(\sigma^2,-i\sigma^1)$
and introducing $\gamma^5\equiv\gamma$ with
$(\gamma)_{\alpha}^{~\beta}=\sigma^3$. 
A vector in $2D$ spinor notation
is now
given by a (symmetric) $\gamma$-traceless pair of indices
$V_{\alpha\beta}=V_{\beta\alpha}$;
$\gamma^{\alpha\beta}V_{\alpha\beta}=0$,
and a $3D$ vector reduces according to
\begin{equation}
V_{\alpha\beta}\rightarrow V_{\alpha\beta}+
\frac{i}{\sqrt{2}}\gamma_{\alpha\beta}v .
\end{equation}
The algebra thus becomes
\begin{equation}
\{\nabla_{\alpha},\nabla_{\beta}\}
=2i\nabla_{\alpha\beta}+2i\gamma_{\alpha\beta}H ,
\end{equation}
defining the $2D$ Yang-Mills field strength 
scalar superfield $H$. Using
projection operators $P_{\pm}\equiv\frac{1}{2}(1\pm\gamma)$ we may
rewrite
this in ``$\pm$-notation'' as
\begin{eqnarray}
\nabla_{\pm}&\equiv&D_{\pm}-i\Gamma_{\pm}, \\
\nabla^{2}_{\pm}&=& \pm i\nabla_{+\!\!\!+ / =}, \\
\{\nabla_{+},\nabla_{-}\}&=&2H,
\end{eqnarray}
which is the form of the $2D$ superspace Yang-Mills algebra we need.

\subsection{2D-supersymmetric actions}
\label{appActions}

The component action of the supersymmetric self-dual massive action is
\begin{eqnarray}
I_{2}(\bar{A},\bar{a},\lambda ,B,b,S,\zeta ,\chi ,V,v)
=\mbox{Tr}\int d^{2}x\left\{\frac{1}{2}m^{2}B_{a}B^{a}
-\frac{1}{2}m^{2}b^{2}
\right. \nonumber \\ \left.
-\frac{1}{2}m\epsilon^{ab}b(\nabla_{a}(V)B_{b}
+\frac{1}{2}B_{a}B_{b})
+\frac{1}{4}m\epsilon^{ab}B_{a}[v,B_{b}]+m\bar{a}F(\bar{A})
\right. \nonumber \\ \left.
+\frac{1}{2}m^{2}S^{2}+m^{2}(i\bar{\zeta}\not\!\nabla(V)\zeta
+\frac{i}{2\sqrt{2}}\bar{\zeta}\gamma[v,\zeta]
-2\bar{\chi}\zeta)
+2m\left(\bar{\lambda}\lambda-\frac{1}{4}\bar{\chi}\chi\right)
\right.\nonumber \\ \left.
-\frac{1}{2}m([S,\bar{\zeta}]
-i[B^{a},\bar{\zeta}]\gamma_{a}
+[b,\bar{\zeta}]\gamma)\left(\eta+\frac{1}{2}\chi\right)
\right. \nonumber \\ \left.
+\frac{1}{4}mF(V)\bar{\zeta}\gamma\zeta
+\frac{i}{2}m\epsilon^{ab}(\nabla_{b}(V)v)
(\bar{\zeta}\gamma_{a}\zeta)
\right. \nonumber \\ \left.
+\frac{i}{8}m\epsilon^{ab}b\nabla_{a}(V)\bar{\zeta}
\gamma_{b}\zeta-\frac{1}{16}m\epsilon^{ab}
B_{a}[v,\bar{\zeta}\gamma_{b}\zeta]
+\frac{1}{8}m\epsilon^{ab}
B_{a}\nabla_{b}(V)\bar{\zeta}
\gamma\zeta
\right. \nonumber \\ \left.
\frac{1}{6}m\bar{\eta}\gamma^{a}[\zeta,\bar{\zeta}\gamma_{a}\zeta]
+\frac{1}{6}m\bar{\eta}\gamma[\zeta,\bar{\zeta}
\gamma\zeta]\right\}+I(V,v).
\end{eqnarray}
After performing the shift $(\bar{A}_{a},\bar{a})=
(A_{a},a)+\frac{1}{2}(B_{a},b)$
the component action transforms into
\begin{eqnarray}
I_{2}(A,a,\lambda ,B,b,S,\zeta ,\chi ,V,v)
=\mbox{Tr}\int d^{2}x\left\{\frac{1}{2}m^{2}B_{a}B^{a}
-\frac{1}{2}m^{2}b^{2}
\right. \nonumber \\ \left.
+\frac{1}{2}mbF(A)+m\epsilon^{ab}B_{a}\nabla_{b}(A)a
\right. \nonumber \\ \left.
+\frac{1}{2}m(a-v)\epsilon^{ab}B_{a}B_{b}
+mb\epsilon^{ab}(A_{a}-V_{a})B_{b}+maF(A)
\right. \nonumber \\ \left.
+\frac{1}{2}m^{2}S^{2}+m^{2}(i\bar{\zeta}\not\!\nabla(V)\zeta
+\frac{i}{2\sqrt{2}}\bar{\zeta}\gamma[v,\zeta]
-2\bar{\chi}\zeta)
+2m\left(\bar{\lambda}\lambda-\frac{1}{4}\bar{\chi}\chi\right)
\right.\nonumber \\ \left.
-\frac{1}{2}m([S,\bar{\zeta}]
-i[B^{a},\bar{\zeta}]\gamma_{a}
+[b,\bar{\zeta}]\gamma)\left(\eta+\frac{1}{2}\chi\right)
\right. \nonumber \\ \left.
+\frac{1}{4}mF(V)\bar{\zeta}\gamma\zeta
+\frac{i}{2}m\epsilon^{ab}(\nabla_{b}(V)v)
(\bar{\zeta}\gamma_{a}\zeta)
\right. \nonumber \\ \left.
+\frac{i}{8}m\epsilon^{ab}b\nabla_{a}(V)\bar{\zeta}
\gamma_{b}\zeta-\frac{1}{16}m\epsilon^{ab}
B_{a}[v,\bar{\zeta}\gamma_{b}\zeta]
+\frac{1}{8}m\epsilon^{ab}
B_{a}\nabla_{b}(V)\bar{\zeta}
\gamma\zeta
\right. \nonumber \\ \left.
\frac{1}{6}m\bar{\eta}\gamma^{a}[\zeta,\bar{\zeta}\gamma_{a}\zeta]
+\frac{1}{6}m\bar{\eta}\gamma[\zeta,\bar{\zeta}
\gamma\zeta]\right\}+I(V,v).
\end{eqnarray}
We may now eliminate $B_{a}$ and $b$ using the equations of motion to
find a
very complicated system that we spare the reader.

\end{document}